# Privacy-Preserving k-Secure Sum Protocol

Rashid Sheikh , Beerendra Kumar
SSSIST, Sehore, INDIA
rashidsheikhmrsc@yahoo.com,
beerucsit@gmail.com

Durgesh Kumar Mishra
Acropolis Institute of Technology and Research,
Indore, INDIA
mishra_research@rediffmail.com

*Abstract*-Secure Multiparty Computation (SMC) allows parties to know the result of cooperative computation while preserving privacy of individual data. Secure sum computation is an important application of SMC. In our proposed protocols parties are allowed to compute the sum while keeping their individual data secret with increased computation complexity for hacking individual data. In this paper the data of individual party is broken into a fixed number of segments. For increasing the complexity we have used the randomization technique with segmentation

*Keywords*- Computation Complexity, Privacy, Random numbers, Secure Multiparty Computation (SMC)

## 1. INTRODUCTION

SMC allows multiple parties to evaluate a common function of their individual data inputs but no party wants to disclose its private data. Many practical situations arise when privacy of data becomes a concern. On the other hand knowing the result of common computation is in their mutual interest. Consider following scenario:

Four brothers living independently want to know the total wealth of family but no brother wants to disclose his individual wealth. All the students in a class want to know the average marks obtained by students but no student is willing to show his marks to others. Certain number of mobile phone companies wants to know the total customers in an area but no company want to disclose its number of customers.

SMC concept was introduced by Yao [1] where he gave a solution to two millionaire's problem. Each of the millionaires wants to know who is richer without disclosing individual wealth. After that the subject has taken many branches like privacy preserving statistical analysis, privacy preserving data mining, privacy preserving SQL query, privacy preserving geometric computation and privacy preserving scientific computation. Privacy preserving secure sum computation is a best and easily understood example of SMC given by Clifton *et al.* [13] which uses random numbers. In our protocol the data of individual party is partitioned into a fixed number of segments. In secure sum protocol one of the parties is selected as the protocol initiator. The protocol initiator passes one segment to next party. The next party adds its segment and passes partial sum to its next party. When summation of first segment of each party is over the protocol initiator adds its next segment to this sum and then passes to the next party .This procedure is repeated until all the segments of the data item are added. Finally the sum in announced by protocol initiator. Proposed protocol is novel in the fact that no random number is used as a key. Therefore it avoids risk of knowing the key in the case when $P_{n-1}$ and $P_i$ maliciously cooperate to know the random number [13]. Since the data is partitioned in to fixed number of segments trying to know the segment of some data will not be fruitful for malicious parties. In this protocol the probability of collecting the data by malicious parties is very low. The number of computing iterations is same as the number of segments each party uses for its data block.

## 2. RELATED WORK

The history of SMC began when Yao [1] proposed his well known Millionaire Problem. This idea was further extended by Goldreich *et al*. [4]. In all these studies theoretical aspect of the concept of SMC was considered. After that few practical problems of SMC were introduced like Private Information Retrieval problem (PIR) [6]. PIR problem uses a client server paradigm in which the client gets $i^{th}$ bit of binary sequence from the server but the server is unaware of that bit. On the other hand the server does not want the client to know the bit sequence. The aim of PIR problem solution is to reduce the communication complexity. Researchers proposed solutions to another specific SMC problem called privacy preserving data mining [2,7,14]. Lindell defined the problem as: two parties, each having a private database, wants to jointly conduct a data mining operation on the union of their two databases. In this, how these two parties accomplish this without disclosing their databases to the other party, or any third party. Agrawal defines the problem as how one party is allowed to conduct data mining operation on a database of other party while preserving the privacy of individual data records. Apart from this many specific SMC problems are studied and solutions were provided by the researchers. Among these are Privacy-Preserving cooperative scientific computations [8], Privacy-Preserving Database Query [9], Privacy-Preserving Intrusion Detection [10], Privacy-Preserving Geometric Computation [11], and Privacy-Preserving statistical analysis [12].

A simple and efficient example of SMC was provided by Clifton *et al.* [13]. He provided secure sum protocol in which all the parties are allowed to compute the sum of their individual data without disclosing it to other parties. In this protocol one of the parties is selected as protocol initiator. The protocol initiator party chooses a random number and then adds this random number to its data. The party then sends this partial sum to the next party. This procedure is repeated until the sum is received back by protocol initiator. Since the random number is known to the protocol initiator party only the actual sum is computed by subtracting the random number and this sum is distributed to all the parties.

### 3. ASSUMPTIONS FOR PROPOSED PROTOCOL
Our protocol preserves privacy of individual as well as provides correct result under following assumptions:-
a. Number of parties must be three or greater.
b. Each party has a computing facility.
c. The communication link between parties is secure.
d. No recipient tells anything about the received partial sum to any other party.
e. All the parties agree on number of segments of a data.
f. Every party follows honestly the protocol for partial sum computation and passes the sum to next party in the
   ring.

### 4. PROPOSED ARCHITECTURE OF *k-Secure Sum Protocol*
As shown in fig 1 all the parties are assumed to be present in a ring. The data block of each party is partitioned into a fixed number of segments. In the shown architecture only four segments are considered. Suppose party $P_0$ is selected as protocol initiator then this party will start the protocol by sending the first segment of its data block. The flow of partial sum will follow a unidirectional ring. The resulting sum is announced by the protocol initiator.

#### 4.1 INFORMAL DESCRIPTION OF *k-Secure Sum Protocol*
Our protocol uses real model of secure multiparty computation where the parties are arranged in a ring. One party among these will be selected as protocol initiator. This protocol initiator will just pass first data segment to next party. If $P_0$ is selected as a protocol initiator it will simply pass first data segment to $P_1$. The party $P_1$ will add the received segment to its first segment and send the partial sum to $P_2$. In general each $P_i$ will send partial sum to $P_{(i+1) \bmod n}$ where $n$ is the number of parties. The protocol initiator initializes a counter say $rc$ (Round Counter) to $k$ where $k$ is the fixed number of segments in each data block. It indicates the number of rounds to be performed for getting the sum of individual data. When one round of computation is completed, the value of $rc$ is decremented by one. When $rc$ reaches zero, the protocol initiator announces the sum. Thus, each party in our protocol must have capability to break the data block into segments and capacity to store each segment.

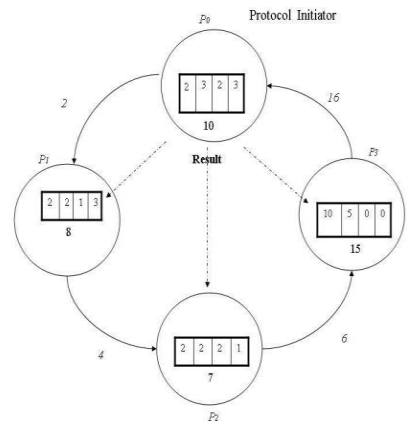

Figure 1: Architecture of k-Secure Sum Protocol

#### 4.2 FORMAL DESCRIPTION OF *k-Secure Sum Protocol*
1. Assume $P_0$, $P_1$, $P_2$,..., $P_{n-1}$ are $n$ parties involved in cooperative $k$-secure sum computation.
2. Assume $k$ is an integer which represents number of segments in each data block.
3. Let $D_0$, $D_1$, $D_2$,..., $D_{n-1}$ are data blocks belonging to $P_0$, $P_1$, $P_2$,..., $P_{n-1}$ respectively.
4. Break $D_i$ into segments $D_{i0}$, $D_{i1}$, ..., $D_{i(k-1)}$ such that $D_i = \sum D_{i,j}$ where $j = 0$ to $k$-1.
5. Assume $rc = k$ and $S_{ij} = 0$,
   /* $S_{ij}$ is partial sum */
6. while $rc \ne 0$
   begin
   for $j = 0$ to $k$-1
   for $i = 0$ to $n$-1
      $P_i$ sends $S_{ij} = D_{i,j} + S_{ij}$ to $P_{(i+1) \bmod n}$
   $rc = rc - 1$
   end
7. $P_0$ announces $S_{i,j}$
8. End of algorithm

### 5. PROPOSED ARCHITECTURE OF *Extended k-Secure Sum Protocol*
As shown in fig 2 the data block is broken into $k$ segments similar to *k-Secure Sum Protocol* but each round of segment summation uses a random number $r_i$. Each round of segment summation uses secure sum protocol.

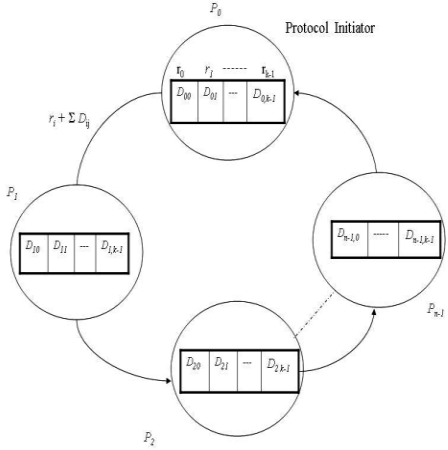

Fig 2 Proposed Architecture of Extended k-Secure Sum Protocol

## 2 FORMAL DESCRIPTION OF THE *Extended k-Secure Sum Protocol*

1. Assume $P_0$, $P_1$, $P_2$,..., $P_{n-1}$ as n parities involved in cooperative *k*-secure sum computation.
2. Each party breaks data block in *k* segments.
3. Let $D_0$, $D_1$, ..., $D_{n-1}$ are data blocks belonging to $P_0$, $P_1$,..., $P_{n-1}$ respectively.
4. Break $D_i$ into segments $D_{i,0}$, $D_{i,1}$,..., $D_{i,k-1}$ such that $D_i = \sum D_{i,j}$ where $j = 0$ to $k$-1.
5. Assume round counter $rc = k$ and partial sum $S_{ij} = 0$.
6. while $rc$!=0
   begin
     for $j = 0$ to $k$-1
     begin
     Select a random number $r_j$
     for $i = 1$ to $n$-1
     begin
     $P_i$ sends $S_{ij} = r_j + D_{i,j} + S_{ij}$ to $P_{(i+1) \mod n}$
     end
     $rc = rc - 1$
     $S_{ij} = S_{ij} - r_j$
     end
   end
7. $P_0$ announces $S_{ij}$.
8. End of algorithm.

## 6. ANALYSIS AND PERFORMANCE

**Case1**: *When all parties are honest*
When all parties are honest the protocol runs in a smooth fashion. Segments are added by all the parties and finally when all rounds of computations are completed the sum is announced by the protocol initiator party. The announced sum is correct and privacy of individual party is preserved. The performance of our protocol in "all-party-honest" case is lower as compared to [13]. As per [13] only one round of computation is needed to get the correct sum. In our protocol since *k* rounds of computations are performed, definitely it is time consuming and costly technique giving more communication and computation complexity to produce the correct sum.

**Case2**: *When the Protocol Initiator becomes malicious*
When the protocol initiator party behaves as a malicious party the segment value transmitted and the partial sum can be incorrect. It will give rise to wrong result. Even in this situation privacy of individual data will be preserved.

**Case3**: *When two neighbor parties turn malicious*
When two parties adjacent to a third party become malicious the middle party will become victim. In this case these two parties communicate with each other to know the data or the segment value of the middle party by taking the difference of partial sum received from corrupt party and that received from middle party. In Clifton's Secure Sum protocol [13] corrupt parties need to perform one computation to know private data of middle party. In our proposed protocol these parties perform one computation to know one segment only. To know all segments of a data block of a party the malicious parties will have to perform *k* such computations. Thus, the computation complexity to break one party data is *k* times as compared to secure sum algorithm [13]. This *k* is number of segment in which each data block is broken. That is the reason why our protocol is named as *k-Secure Sum Protocol*. In *Extended k-Secure sum Protocol* the corrupt parties will have to perform 2*k* computations because one additional computation will be needed to know the random number. The probabilistic analysis shows that as number of parties increases the probability of a node becoming victim decreases. This is depicted in *figure 1* for *Secure Sum Protocol* as given by Clifton *et al.* [13]. In *k-Secure Sum Protocol* the same probability can further be reduced by increasing the value of *k* as depicted in fig 3.

Let n be number of parties. In Secure Sum Protocol [13] the probability of one party becoming victim by any two neighbors is given by:

$P_1 = n / {}^nC_2$

$P_1 = 2/n-1$ where, $n \geq 3$         (1)

In our *k*-Secure Sum Protocol using k segments of a data block the probability is given by-

$$P_{1k} = (2/n-1)^k \qquad (2)$$

Note that since *2/n-1* is less than 1, raising the power *k* causes probability to decrease. Thus, the probability analysis of the *k*-Secure Sum Protocol indicates that our protocol is more secure than Secure Sum Protocol given by Clifton *et al*. Extending the protocol by using one random number for each round of computation further increases computation complexity and thus makes it more difficult to know the data of victim party. In fig 3 the case *k*=1 is the probability curve for Clifton's Secure Sum Protocol and other curves show the probability of our protocols.

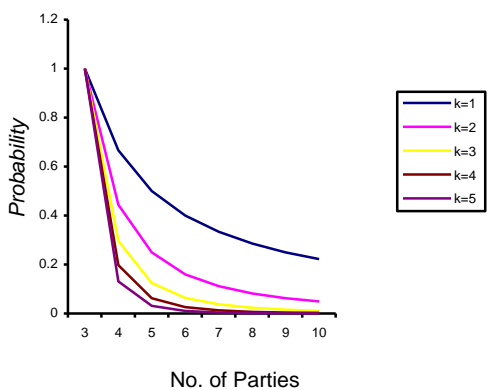

Fig.3: Probability of party becoming victim in k-secure sum protocol

## 7. CONCLUSION AND FUTURE SCOPE

Our *k-Secure Sum Algorithm* and *Extended k-Secure Sum Algorithm* are used to get the sum of private data belonging to all parties providing lower probability of data leakage. The probability analysis shows that this is an appreciable improvement over previous protocol. It provides excellent security when number of segments is sufficiently large. If all parties work honestly the protocol provide correct result maintaining the privacy of individual data. The *k-Secure Sum Protocol* improves complexity *k* times as compared to previous protocols. *Extended k–Secure Sum Protocol* provides more than *k* computations to malicious parties for breaking data of a victim party. Further effort can be made to make the protocol having more computation complexity. One way is that each party keeps one segment with it and *k*-1 segments are distributed to other parties. After this distribution, the segments kept by a party will not belong to the data block of a single party. The *Secure* Sum *Algorithm*, *k-Secure Sum Algorithm* and *Extended k-Secure Sum Algorithm* may now be applied.

*Conference*, New Orleans, Louisiana, USA, pages 102-110, Dec. 10-14 2001.

## AUTHORS PROFILE

*Rashid Sheikh*
Ph. +91 9826024087
Email: rashidsheikhmrsc@yahoo.com

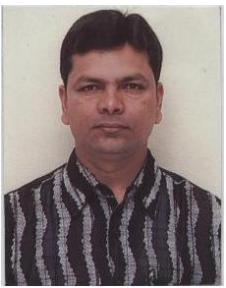

Rashid sheikh has received his Bachelor of Engineering degree in Electronics and Telecommunication Engineering from Shri Govindram Seksaria Institute of Technology and Science, Indore, M.P., India in 1994. He has 15 years of teaching experience. His subjects of interest include Computer Architecture, Computer Networking, Electrical Circuit analysis, Digital Computer Electronics, Operating Systems and Assembly Language Programming. Presently he is pursuing M. Tech. (Computer Science and Engineering) at SSSIST, Sehore, M.P., India. He has published three research papers in National Conferences. His research areas are Secure Multiparty Computation and Mobile Ad hoc Networks. He is the author of ten books on Computer Organization and Architecture.

*Durgesh Kumar Mishra, PhD*
Ph - +91 9826047547, +91-731-4730038
Email: durgeshmishra@ieee.org

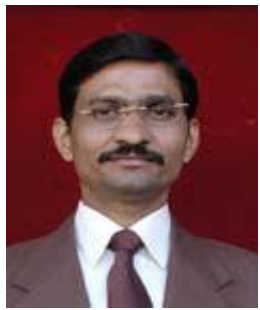

Durgesh Kumar Mishra has received his M.Tech. degree in Computer Science from DAVV, Indore in 1994 and PhD degree in Computer Engineering in 2008. Presently he is working as Professor (CSE) and Dean (R&D) in Acropolis Institute of Technology and Research, Indore, MP, India. He is having around 20 Yrs of teaching experience and more than 5 Yrs of research experience. He has completed his research work with Dr. M. Chandwani, Director, IET-DAVV Indore, MP, India on Secure Multi-Party Computation. He has published more than 60 papers in refereed International/National Journals and Conferences including IEEE and ACM. He is a senior member of IEEE and Secretary of IEEE MP-Subsection under the Bombay Section, India. Dr. Mishra has delivered tutorials in IEEE International conferences in India as well as other countries. He is programme committee member of several International conferences. He visited and delivered invited talks in Taiwan, Bangladesh, USA, UK, etc. on Secure Multi-Party Computation of Information Security. He is an author of one book. He is reviewer of three International Journals of Information Security. He is a Chief Editor of *Journal of Technology and Engineering Sciences*. He has been a consultant to industries and Government organizations like Sales Tax and Labor Department of Government of Madhya Pradesh, India.